# Nanoscale control of exchange bias with BiFeO$_3$ thin films

*Lane W. Martin*[*†‡], *Ying-Hao Chu*[†‡**], *Mikel B. Holcomb*[†§], *Mark Huijben*[§**], *Shu-Jen Han*[∥], *Donkoun Lee*[∥], *Shan X. Wang*[∥], *and R. Ramesh*[†‡§]

Materials Science Division, Lawrence Berkeley National Laboratory, Berkeley, CA 94720 USA;

Department of Materials Science and Engineering, University of California, Berkeley, CA 94720 USA;

Department of Physics, University of California, Berkeley, CA 94720 USA; Department of Materials Science and Engineering, Stanford University, Stanford, CA 94305 USA

*lwmartin@lbl.gov

*Corresponding author: lwmartin@lbl.gov

[†]Materials Science Division, Lawrence Berkeley National Laboratory, Berkeley

[‡]Department of Materials Science and Engineering, University of California, Berkeley

[§]Department of Physics, University of California, Berkeley, CA 94720 USA

[∥]Department of Materials Science and Engineering, Stanford University, Stanford, CA 94305 USA

**Currently at Department of Materials Science and Engineering, National Chiao Tung Unviersity, HsinChu, 30010, Taiwan

We demonstrate a direct correlation between the domain structure of multiferroic $BiFeO_3$ thin films and exchange bias of $Co_{0.9}Fe_{0.1}/BiFeO_3$ heterostructures. Two distinct types of interactions – an enhancement of the coercive field (*exchange enhancement*) and an enhancement of the coercive field combined with large shifts of the hysteresis loop (*exchange bias*) – have been observed in these heterostructures, which depend directly on the type and crystallography of the nanoscale (~2 nm) domain walls in the BFO film. We show that the magnitude of the exchange bias interaction scales with the length of 109° ferroelectric domain walls in the $BiFeO_3$ thin films which have been probed via piezoresponse force microscopy and x-ray magnetic circular dichroism.

Multiferroics, or materials that simultaneously show some magnetic and ferroelectric order, such as BiMnO$_3$[1] and BiFeO$_3$[2], have piqued the interest of researchers worldwide with the promise of coupling between magnetic and electric order parameters.[3,4] Over the last few years, much has been learned about the underlying interactions in these intrinsic multiferroics and how one can control the properties of these materials. BiFeO$_3$ (BFO), a multiferroic with a ferroelectric Curie temperature of ~820°C and an antiferromagnetic Neél temperature of ~370°C,[5,6] has been the focus of many papers and much has been learned about how to control the ferroelectric domain structure,[7,8] the domain switching mechanisms,[9] and, in turn, the coupling between ferroelectric and magnetic order parameters.[10,11,12]

At the same time, great advances in exchange anisotropy have occurred since the first discovery of this phenomenon in 1956.[13] Exchange anisotropy or bias (EB) describes the phenomena associated with the exchange anisotropy created at the interface between an antiferromagnet and a ferromagnet.[14] Heterostructures based on multiferroic materials, including YMnO$_3$[15,16] and BFO,[17,18] have demonstrated strong static exchange interactions . To this point, however, a robust, room temperature exchange coupled system that is electrically tunable[19] has yet to be experimentally demonstrated. In this letter we report a fundamental and direct correlation between nanoscale features – specifically ferroelectric domain walls – in BFO and the nature of the exchange interaction between the ferromagnet Co$_{0.9}$Fe$_{0.1}$ (CoFe) and the multiferroic, antiferromaget BFO.

Heterostructures of Ta (5 nm) / CoFe (2.5 nm) / BFO (50-200 nm) were grown on SrTiO$_3$ (STO) and SrRuO$_3$ (SRO) / STO (001) oriented substrates using pulsed laser deposition, details of which have been reported elsewhere.[7,8] The key parameter of relevance to this work is the deposition rate for the growth of the BFO layer, which was varied from ~0.1-0.3 Å/sec to 1-2 Å/sec. Following the deposition of the BFO/SRO, the films were cooled in 1 Atm of oxygen to room temperature. The surface structure and underlying FE domain structure were then analyzed using a combination of atomic force microscopy (AFM) and piezoresponse force microscopy (PFM). AFM studies of our BFO heterostructures reveal root-mean-square roughness values of ~0.60 nm regardless of the underlying ferroelectric domain structure. Transmission electron microscopy and X-ray diffraction analysis reveal single phase, fully epitaxial BFO films. The samples were then transported to an ion beam sputtering system with a base pressure of ~5E-9 Torr where the ferromagnetic CoFe alloy and capping layer films were grown. In traditional EB systems, the effect is only observed upon cooling the system through the Neél temperature ($T_N$) of the antiferromagnet in the presence of an external applied field. In our system, heating to above $T_N \approx 370°C$ resulted in inter-diffusion of the layers and oxidation of the CoFe films. We have therefore circumvented this issue by growing the CoFe films in an applied field, $H_{Growth} = 200$ Oe, so as to induce a uniaxial anisotropy. Magnetic properties were measured using a SHB Instruments, Inc. Loop Tracer and Quantum Design SQUID magnetometer. X-ray magnetic circular dichroism (XMCD) measurements were carried out at beamline 4.0.1 of the Advanced Light Source at Lawrence Berkeley National Laboratory.

Two distinctly different types of room temperature magnetic responses are observed in these heterostructures [Figure 1(a) and (b)]. One type of sample exhibits small negative shifts of the hysteresis loop (typical EB field ($|H_{EB}|$) ~ 2.5-10 Oe), along with a significant enhancement of the coercive field ($H_C$) ~ 30-70 Oe as compared to that of CoFe grown directly on STO (001) ($H_C^{CoFe}$ ~ 5-7 Oe) [Figure 1(a)]. This same subset of heterostructures also exhibits a hard axis of magnetization when measured perpendicular to $H_{Growth}$, that arises from the uniaxial anisotropy induced in the CoFe during growth. It is also noted that the strong vertical shift between the ferromagnet half-loops is likely due to enhanced spin viscosity during spin rotation in the antiferromagnetic layer. In this manuscript we will refer to heterostructures exhibiting this behavior as possessing an *exchange enhancement* (i.e., an enhanced spin viscosity or spin drag effect)[14] interaction. In contrast, other heterostructures have been found to exhibit magnetic properties such as those shown in Figure 1(b), namely, a strong negative $H_{EB}$ (typical $|H_{EB}|$ ~ 50–150 Oe) when measured parallel to $H_{Growth}$ as well as enhancements of $H_C$ (typical $H_C$ ~ 15-80 Oe). We observe the opposite shift of the hysteresis loop when measuring antiparallel to $H_{Growth}$ and the formation of a hard axis when measuring perpendicular to $H_{Growth}$ [Figure 1(b)], thus confirming the behavior to be an *exchange bias* interaction.

Since the net magnetization of an antiferromagnet is zero it has been hypothesized that the in-plane symmetry of the antiferromagnet must be broken to account for the EB. Over the years a number of factors such as interfacial roughness,[20] grain/domain size,[21,22] non-magnetic defect sites,[23,24] or a combination of these factors have been proposed to explain the formation of the pinned, uncompensated surface spins required for EB. We demonstrate that the nature and density of specific types of domain walls in the BFO layer is the most critical element in determining EB in this system. To understand this, we recall that in rhombohedral ferroelectrics, such as BFO, there exist three types of domain walls, namely those that separate domains with 71°, 109° and 180° differences in polarization direction. The 71° domain walls form on {101}-type crystallographic planes, which also correspond to a mirror-plane in the rhombohedral structure of BFO, while 109° domain walls form on {001}-type planes.[25] In the case of BFO thin films, it has been demonstrated that careful control of the deposition parameters, such as the film growth rate allows us to reliably obtain a spectrum of domain architectures.[7] At low growth rates (~0.1-0.3 Å/s) a mixture of two orthogonal, stripe-like polarization variants [Figure 1(c)] is observed while at high growth rates (~1-2 Å/s) a highly disordered, mosaic-like domain architecture is observed [Figure 1(d)]. Detailed analyses, combining both in-plane [Figure 1(c) and (d)] and out-of-plane [inset in Figure 1(c) and (d)] PFM images, allows for the determination of the underlying ferroelectric domain structure. The results of these analyses are summarized in Figure 1(e) and (f) in which all three types of domain walls we have been identified. Further details of our domain wall analyses (i.e., 71° vs. 180° vs. 109°) are described elsewhere.[9] This analysis shows that the stripe-like structures [Figure 1(c) and (e)] correspond to arrays of 71° domain walls while the mosaic-like architecture [Figure 1(d) and (f)] is comprised of a mixture of all possible domain wall types, particularly, large fractions of 109° domain walls and smaller fractions of 71° and 180° walls.

After extensive study, a critical correlation emerges connecting the *exchange enhancement* [Figure 1(a)] and *exchange bias* [Figure 1(b)] interactions to the underlying ferroelectric domain structure of the BFO film. Specifically BFO heterostructures with stripe-like [Figure 1(c)] and mosaic-like [Figure 1(d)] ferroelectric domain structures give rise to *exchange enhancement* and *exchange bias* properties, respectively. It is important to reiterate that regardless of the underlying domain structure of the BFO film, strong coupling between the CoFe and BFO is observed and is manifested as an enhancement of $H_C$ for all heterostructures measured. On the contrary, |$H_{EB}$| is specifically related to the domain complexity of the underlying BFO film and the different nanoscale features found in these films.

There are two important variables of relevance to interpret our data. The first is the fraction (and in turn length) of the different types of ferroelectric domain walls; the second is the fraction of the film surface that is comprised of these different types of ferroelectric domain walls. Standard image analyses[26] of the in-plane and out-of-plane domain images were used to extract quantitative information about the average domain sizes. Using a reasonable ferroelectric domain wall width of 2 nm,[27,28] (we note also that the 109° wall width has been measured directly from atomic resolution TEM images of our samples to be ~2 nm)[29] we calculated the fraction of the surface that is made up of domain walls as well as the fraction of the different types of ferroelectric domain walls (i.e., 71°, 109°, or 180°) in each sample. This analysis was completed on a large set of mosaic-like and stripe-like BFO films and reveals an average 109° wall fraction of ~40-50% in the mosaic-like samples and ~5-10% in stripe-like BFO films. We also note that the fraction of the surface that is made up of such walls scales inversely with the average domain size (as discussed in Figure 2).

We can estimate the coupling strength at the ferromagnet-antiferromaget interface using a simple Heisenberg-like model of exchange interaction where $H_{EB}$ can be written as:

$$H_{EB} = \frac{\sigma}{M_{FM} t_{FM}} = \frac{J_{ex} S_{AFM} S_{FM}}{a_{AFM}^2 M_{FM} t_{FM}} \tag{1}$$

where σ is the unidirectional interfacial energy, $J_{ex}$ is the Heisenberg-like interface exchange energy (~5 meV)[30,31], $S_{AFM}$ and $S_{FM}$ are the sublattice magnetic moment of the BFO and atomic moment of CoFe respectively, $a_{AFM}$ is the lattice parameter of BFO (3.96Å), $M_{FM}$ is the magnetization of CoFe (1591.1 emu/cc), and $t_{FM}$ is the thickness of the CoFe (2.5 nm). If the entire surface is magnetically uncompensated (and therefore contributing to EB) we expect $|H_{EB}|$ ~16 kOe for the CoFe/BFO interface. This estimation, however, fails to recognize the nanoscale origin of the exchange interactions at the interface and hence it greatly overestimates the coupling strengths by assuming uniform coupling over an ideally smooth, uncompensated surface. Furthermore, from an atomic moment picture of the G-type antiferromagnetic structure of BFO, the (001) surface is fully compensated and therefore should give rise to no EB. Hence, we propose that surface magnetic heterogeneities, such as the nanoscale domain walls, are responsible for the EB in our system. More specifically, we propose that the coupling leading to EB primarily arises from uncompensated spins that occur at certain types of nanoscale domain walls in the BFO film as elaborated below.

Substantiation for this hypothesis comes from calculations based on the detailed PFM analyses of domain structures in BFO. For example, at an average domain size of 200 nm, ~2% of the entire surface of both a mosaic- and stripe-like BFO film would be made up of domain walls. Therefore, if we scale our estimates of $|H_{EB}|$ assuming that *all* ferroelectric domain walls contribute pinned, uncompensated spins to the EB interaction we would expect both mosaic-like and stripe-like BFO films to give rise to $|H_{EB}|$ ~300-350 Oe. This estimate is again much higher than experimentally measured for all heterostructures. On the other hand, if we postulate that pinned, uncompensated spins occur *only* at 109° domain walls – the mosaic- and stripe-like BFO films have the biggest difference in the density and length of such ferroelectric domain walls – the resulting $|H_{EB}|$ is ~144 Oe and ~8 Oe for mosaic-like and stripe-like BFO films, respectively. These values are reasonably consistent with experimentally measured values of $|H_{EB}|$ for heterostructures based on stripe-like (typical $|H_{EB}|$ ~ 2.5-10 Oe) and mosaic-like (typical $|H_{EB}|$ ~ 50-150 Oe) BFO films. This leads us to believe that the coupling leading to EB occurs primarily at the surface intersection of 109° ferroelectric domain walls in the BFO layer. Theoretical studies based on symmetry analyses have indeed shown that such nanoscale domain walls in antiferromagnetic, magnetoelectric crystals can carry a spontaneous magnetization.[32]

If the above hypothesis is correct, then clearly the areal density of such 109° domain walls should have a marked influence on the magnitude of the EB. Indeed, previous studies of EB systems have noted the importance of domain complexity and have suggested an inverse relationship between antiferromagnetic domain size and $H_{EB}$.[18,20,33] In order to probe this in the CoFe/BFO system, we have measured $|H_{EB}|$ in heterostructures created from BFO films controlled to have a wide range of domain sizes [Figure 2]. In the case of the samples with stripe-like domains structures, the domain sizes were systematically varied by changing the thickness of the BFO layer; in the case of the mosaic samples, the domain sizes can be varied by many different pathways, including the growth rate, the substrate miscut, the SRO growth mode, etc. Much like the room temperature measurements in Figure 1(a) and (b) there are two distinct behaviors. The heterostructures based on mosaic-like BFO exhibit a monotonic relationship between $|H_{EB}|$ and the inverse of domain size (a line is drawn in Figure 2 to aid the eye). On average, such heterostructures are found to have a much larger $|H_{EB}|$ compared to heterostructures based on the stripe-like BFO films. Furthermore, heterostructures based on stripe-like BFO films exhibit consistently smaller or negligible $|H_{EB}|$ as well as little change in $|H_{EB}|$ as a function of domain size. We have additionally plotted $|H_{EB}|$ for the same set of samples as a function of the total length of 109° domain walls / sample surface area in 5 x 5mm samples (data for both mosaic- and stripe-like heterostructures are shown) [Figure 2(b)]. The length of 109° domain walls was calculated from the in-plane and out-of-plane PFM images. Note, again, that the magnitude of $|H_{EB}|$ increases *monotonically* with the total length of 109° ferroelectric domain wall nano-features.

Temperature dependent magnetic measurements from 5-300K [Figure 3] provide complementary insight into the nature and mechanism of coupling in these heterostructures. These measurements reveal a number of interesting aspects. First, regardless of the underlying BFO domain structure, all samples show a marked increase in $H_C$ with decreasing temperature. We have included temperature dependent properties of a Ta/CoFe/STO (001) film grown under the same conditions for comparison [data in green symbols in Figure 3]. From this data we can see that without the BFO layer, the CoFe layer has a very small change in $H_C$ as temperature is decreased; this points to a strong temperature dependent interaction between CoFe and BFO. The $|H_{EB}|$, however, exhibits very little temperature dependence and remains essentially constant for heterostructures grown on both mosaic- ($H_{EB}$ ~ 50 Oe) and stripe-like BFO films ($H_{EB}$ ~ 0 Oe).

The temperature dependent data gives two very important insights. The first is that the temperature dependent increase in $H_C$ observed for all CoFe/BFO heterostructures is larger than that expected for the CoFe film alone. This implies that the exchange interaction responsible for the enhancement of $H_C$ grows stronger at lower temperatures and is the same regardless of the underlying BFO structure. This suggests that it is related to the macroscopic spin configuration of the (001) BFO surface, which is common to both the stripe- and mosaic-like films. The second finding is that the mechanism which gives rise to EB occurs *only* in mosaic-like BFO films, regardless of the temperature. It should be noted, that both 71° and 109° are ferroelastic domain walls and thus if both were actively serving as pinning sites should give rise to strong EB interaction in both mosaic- and stripe-like BFO structures. The temperature dependent data, along with the data in Figure 2, however, reveals that this is not the case and that even at low temperatures the interfacial spins in the stripe-like BFO structures do not become active pinning centers. The role of uncompensated spins in the evolution of EB has been shown to be important in numerous systems including Permalloy/CoO,[21] Co/FeMn,[34] Co/IrMn,[35] and Co/NiO.[36] More recently it has also been reported that the fraction of pinned, uncompensated spins responsible for EB can be just a small fraction (a few percent) of the entire surface spins.[31] Ohldag, *et al.* conjecture that the origin of the pinned spins could be those spins found at grain boundaries (and domain boundaries) in the antiferromagnet. Our experimental observations are similar to this model in that only a few percent of the surface spins are pinned in the BFO and we believe they occur at specific nanoscale features.

To quantitatively estimate the differences in surface spin structure in stripe- and mosaic-like BFO films we have used XMCD measurement completed at the Fe *L*-edge in total electron yield configuration with a beam area of ~0.01 mm$^2$. The difference in the absorption spectrum for right- and left-circularly polarized light is a measure of uncompensated spins in the material; this asymmetry is thus proportional to the magnetic moment. Figure 4(a) and (b) are x-ray absorption data collected from BFO films with stripe- and mosaic-like domain structures, respectively. The stripe-like BFO films exhibits essentially no measurable asymmetry [Figure 4(c)]; in contrast, the mosaic-like BFO films consistently exhibit normalized asymmetries of between 0.5-1% (at zero applied field) [Figure 4(d)] (in each case, 4 films have been measured and the data shown in Figure4 is a representative data set). The presence of circular dichroism in the mosaic films, even at zero applied field, strongly supports the possibility of the existence of correlated spins (for example, at 109° domain walls) that lead to the EB interactions with the CoFe layer. From these measurements we can also extract an average magnetic moment for the spins in the probed area. We have completed dichroism measurements on $Fe_3O_4$ ($M_S$ = 477.465 emu/cc at room temperature[37]), which typically exhibits an XMCD signal of ~14%,[38] which enables us to estimate the relative moment of the mosaic-like BFO films. Based on the experimentally observed XMCD of 0.5-1.0% for the mosaic-like samples, we estimate a magnetic moment in the range of 17-34 emu/cc. This translates to a magnetic moment of 0.12-0.24 $\mu_B$/Fe. We believe that these values are quite reasonable. Finally, we note that macroscopic SQUID measurements of the same mosaic-like samples, yield saturation moment values in the range of 18-25 emu/cc, thus further validating our findings. Substituting this value into Eq. (1) we can estimate the magnitude of |$H_{EB}$| to be between 100-300 Oe over the range of XMCD measured in our samples. Once again, this order of magnitude estimate is consistent with the EB shifts measured in our heterostructures and points to the connection between the nanoscale domain structure and EB properties in this system.

In summary, our results indicate that there are two major interactions occurring in these heterostructures. One is a surface coupling between the spins in the antiferromaget and the ferromagnet. Detailed magnetic measurements [Figure 1] indicate that this interaction results in very little unidirectional pinning of the ferromagnetic layer and manifests itself as an enhancement of $H_C$. This is the case for the samples showing only *exchange enhancement*. The second interaction, the effect most important for the EB observed here, appears to be a coupling phenomenon at or near the few nanometers where the 109° domain walls in BFO intersect the film surface. It was found that the magnitude of this EB interaction can be tailored by engineering the underlying domain structure of the BFO film thus presenting the ability to gain nanoscale control of EB interactions in an exciting multiferroic based system.

**Acknowledgement.** This work was supported by the Director, Office of Science, Office of Basic Energy Sciences, Materials Sciences and Engineering Division, of the U.S. Department of Energy under Contract No. DE-AC02-05CH11231. The authors would also like to acknowledge the support of the Western Institute of Nanoelectronics and the staff and facilities at the National Center for Electron Microscopy.

**Figure Captions**

**Figure 1.** Room temperature magnetic properties for heterostructures exhibiting *exchange enhancement* [(a)] and *exchange bias* [(b)] properties. (c) and (d) show in-plane and out-of-plane (inset) PFM contrast for typical BFO films that exhibit *exchange enhancement* and *exchange bias*, respectively. Detailed domain wall analysis for (e) stripe-like and (f) mosaic-like BFO films.

**Figure 2.** (a) Dependence of exchange bias field on domain size for CoFe/BFO heterostructures grown on mosaic-like (blue) and stripe-like (red) BFO films. (b) Exchange bias field of the same samples here graphed as a function of the total length of 109° domain walls / sample surface area in 5 x 5mm samples.

**Figure 3.** Temperature dependent magnetization data for CoFe/BFO heterostructures grown on BFO films with stripe-like (circles) and mosaic-like (squares) domain structures. Also included are temperature dependent data for CoFe/STO(001) for comparison.

**Figure 4.** X-ray magnetic circular dichroism measurements on BFO films exhibiting (a) stripe-like and (b) mosaic-like domain structures. (c) and (d) are the respective asymmetry values for each measurement.

Figure 1

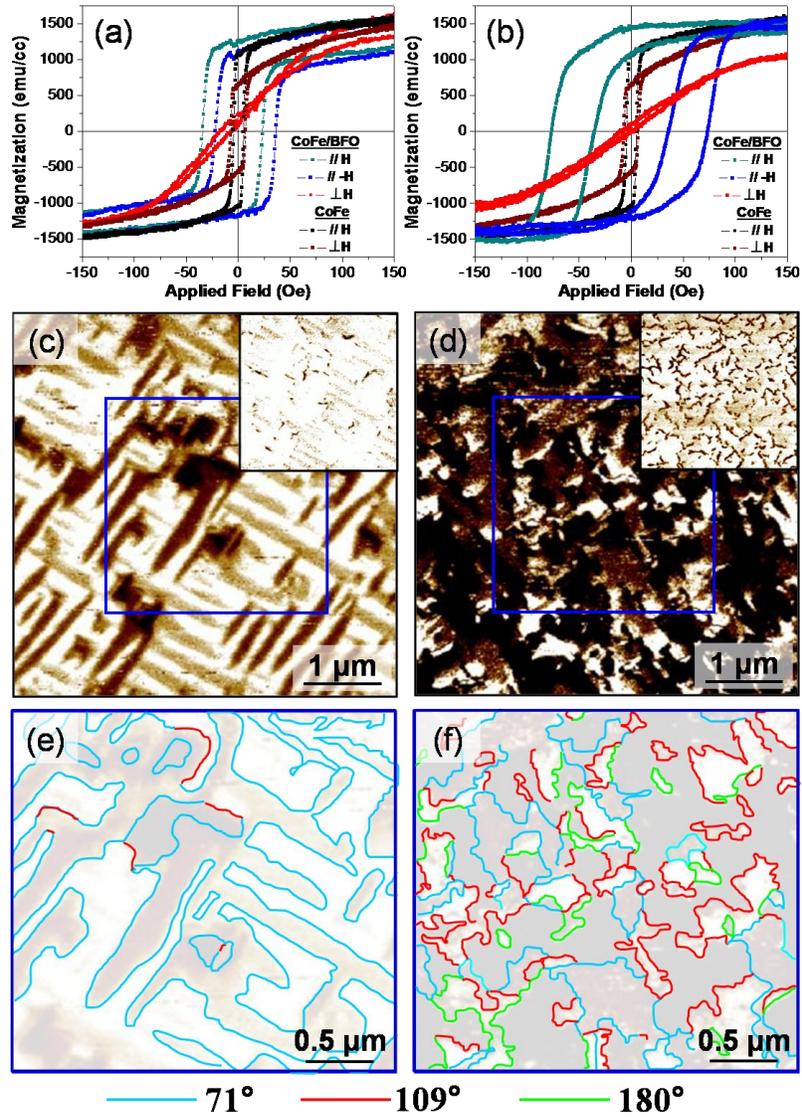

Figure 2

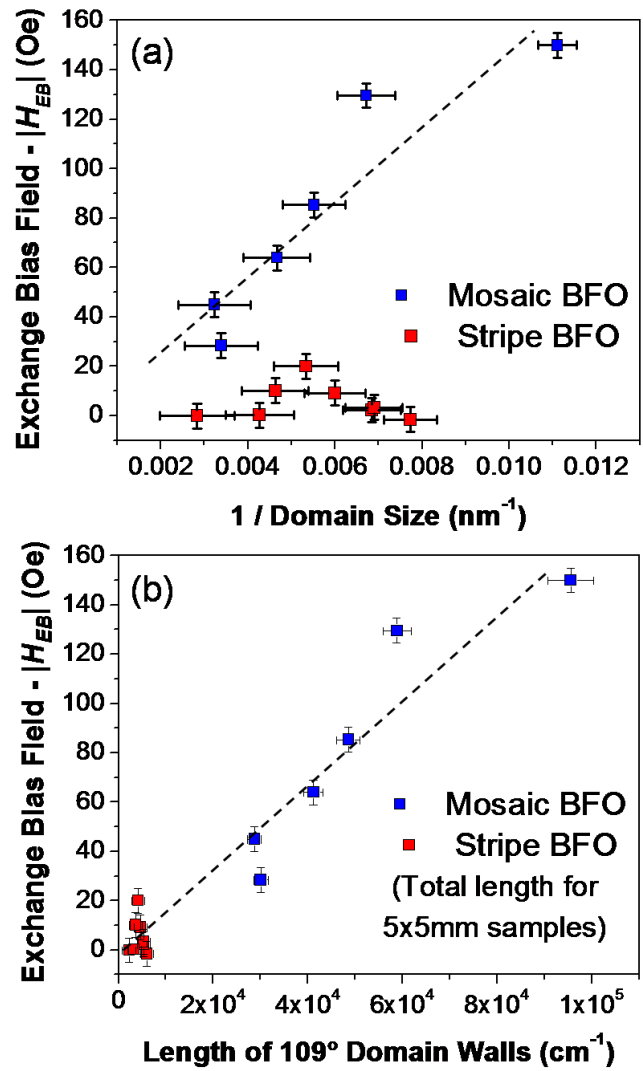

Figure 3

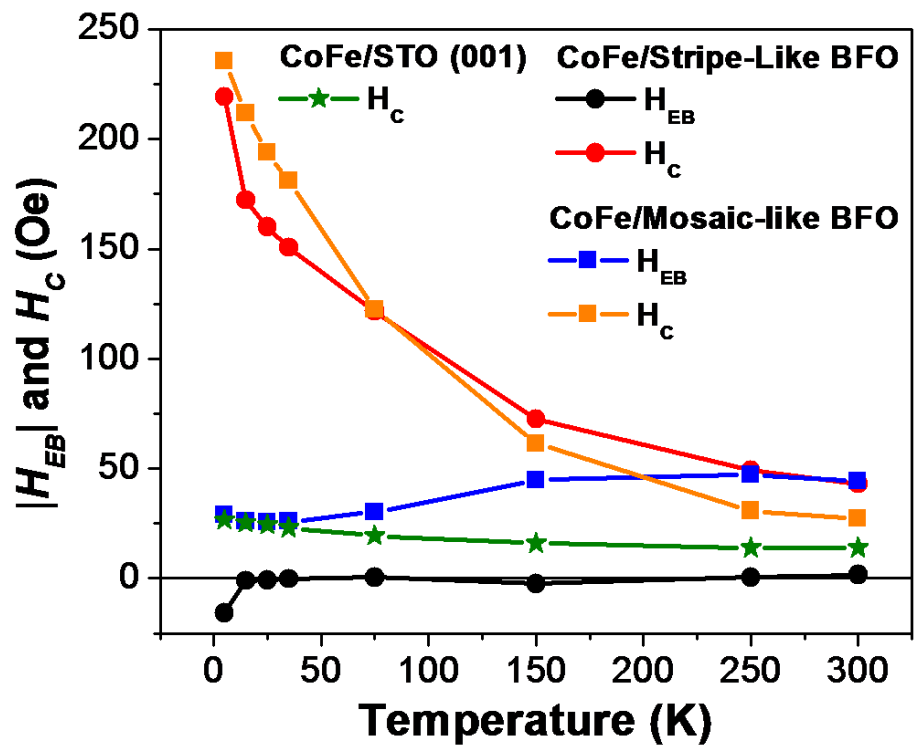

Figure 4

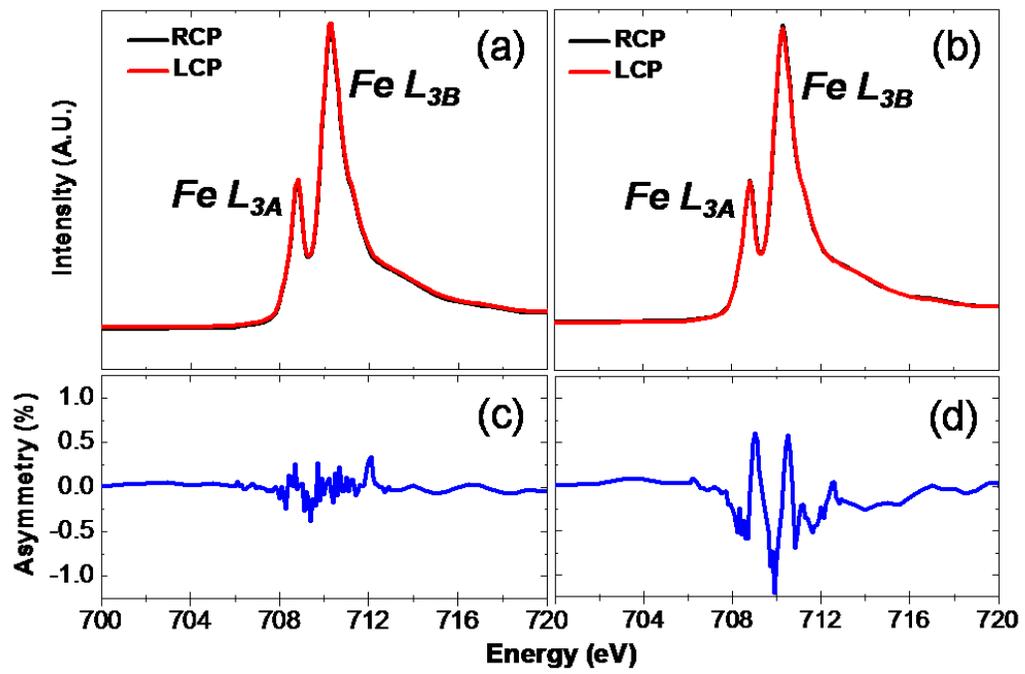